\documentstyle[aps,epsfig]{revtex}
\newcommand\T{\rule{0pt}{2.6ex}}      
\newcommand\B{\rule[-1.2ex]{0pt}{0pt}}
\newcommand\TT{\rule{0pt}{4ex}}      
\newcommand\BB{\rule[-2.6ex]{0pt}{0pt}}
\begin{document}
\draft
\preprint{Preprint number}
\title{Spatial and energy distribution of muons in $\gamma$-induced 
air showers}
\author{A. Fass\`o}
\address{Radiation Physics Department, Stanford Linear Accelerator Center, 
Stanford, CA 94309}
\author{J. Poirier}
\address{University of Notre Dame, Center for Astrophysics, Physics Department, 
Notre Dame, IN 46556}
\date{\today}
\maketitle
\begin{abstract}
The {\sc FLUKA} Monte Carlo program is used to calculate the effects 
of hadroproduction by primary gamma rays incident 
upon the earth's atmosphere; 
for the results presented in this paper, only primary angles at 0 degrees
from zenith are considered.  
The {\sc FLUKA} code 
is believed to be quite accurate in reproducing experimental 
photon hadroproduction 
data in the 1~GeV to 10~TeV energy range studied.  The charged 
pions which are so produced can decay to muons with sufficient 
energy to reach ground level.  The number of these muons and their 
radial and energy distribution are studied for incident gamma 
ray energies from 1~GeV to 10~TeV.  The number of these muons 
is not negligible; they can, in certain circumstances, be 
used to study potential sources of gamma rays like 
gamma ray bursts. It is found, for example, that a 10~TeV incident primary 
gamma ray produces, on average, 3.4 muons which reach ground level; 
the gamma ray energy which produces the maximum number of muons at 
ground level depends on the spectral index of the primary gamma spectrum, 
a constant which describes how the primary gamma flux rises with 
decreasing primary energy.  An example: for a differential spectral 
index of 2.7, there is a broad maximum number of muons coming from 
$\approx 30$~GeV primary gamma ray energy.
\end{abstract}
\pacs{PAC numbers: 95.75.Pq,98.70.Sa,98.70.Rz,13.60.Le}

\section{Introduction}
\label{intro.sec}
Primary cosmic rays consist mainly of an isotropic flux of charged
particles (primarily protons and nuclei).  
Before reaching the earth, galactic
magnetic fields deflect their paths so information about the
angular position of their source is lost.  On the contrary, neutral
particles such as gammas and neutrinos (neutrons decay before reaching
the earth) can be directly used to locate 
the angular position of their origin.
Therefore gamma astronomy is 
important in the 
study of well localized exotic astrophysical objects. 
For example, recent reviews are \cite{Ong98,Hof99}.

Cosmic gamma rays can be detected directly only by 
satellite or balloon experiments located 
essentially outside the earth's atmosphere.
The detector size and weight sets 
limits to the sensitivity and energy range which these experiments 
can cover which, at present, 
extends to several tens of GeV. 
Higher energies are better investigated by means of ground-based
experiments which 
sample the numerous secondary particles
produced by the high energy primary photons when they interact in the 
atmosphere.

Photons constitute only a small fraction of primary cosmic rays. 
Hadronic showers at ground level are very similar to electromagnetic 
showers because, at each generation of hadronic pion production, 
about one-third of the pions are $\pi^0$ which immediately decay to gamma rays 
which then initiate electromagnetic sub-showers.  
By the time the hadronically initiated shower reaches ground level it is 
essentially all electromagnetic because of the many generations, 
each feeding one-third of their energy into the electromagnetic sector.  
Therefore, gamma showers at ground level 
are qualitatively similar to those produced
by protons and nuclei \cite{Gai90}, and 
experiments based on earth which search for gamma primaries 
must subtract a large background due to the more numerous cosmic protons
and nuclei. 
Various methods have been developed to discriminate
between the two types of showers \cite{Ong98,Hof99,Hil96,Sch96}. One of
the most commonly used techniques is based on their different muon
content \cite{Gai91}.

Muons are created in the atmosphere mainly as decay products of charged
mesons, which are abundant in hadronic cascades. Pions and kaons can
also be produced in the nuclear interactions of high energy primary gamma rays
which then initiate secondary hadron showers.
The photon cross section for hadronic 
interactions is about two orders of magnitude smaller than that 
for producing electron pairs.  
This low hadronic cross section for gamma rays has been used 
as a signature for showers initiated by hadrons.    

However, muons are present in gamma showers too,
albeit in smaller numbers than in a hadronically initiated shower 
of the same primary energy.  
The first estimates confirmed the ``muon-poor'' characteristics of
gamma showers \cite{Kar63,McC79}.
However, in 1983, Samorski and Stamm \cite{Samorski} 
reported the results of an extensive air shower 
experiment at Kiel claiming an excess of events 
with energies above 2000~TeV centered at the 
angular location of Cygnus X-3.  
They also reported a non-deficiency 
of muons in these data, seemingly inconsistent with a primary gamma ray 
hypothesis.  
In order to evaluate and discuss these data, many
authors performed analytical or Monte Carlo calculations of the muon
flux produced in gamma showers 
\cite{Kud85,Edw85,Pro85,Sta85a,Sta85b,Sta85c,Hal86,Sta86,Ber88,Dre89,Cha90,Kry91}. 
Most of these calculations were
one-dimensional and all of them referred to gamma energies 
much larger than 10~TeV, appropriate to the Kiel experiment.  The energy realm 
of these calculations was considerably larger than the 1~GeV to 10~TeV region 
considered in this paper.
Therefore, their stress was mostly on 
sources of muons at very high energies (Bethe-Heitler
$\mu$-pair production, decay of charmed particles), 
with less emphases on muon decay,
energy loss, and radial distribution which become important in the 
energy range considered in this paper.  

More recently, the same calculation techniques have been refined and
used to optimize current experiments. In addition to an improvement
of the adopted cross sections \cite{Bha97}, muon energy loss and decay
have been added \cite{Hal97,Alv99}, in order to better estimate the 
sensitivity of modern experiments having a lower muon energy threshold.
Several experiments, GRAND \cite{GRAND} and Milagro \cite{Atk99} for example, 
have the capability of measuring 
the angles of muons at ground level with high statistics.  
With the advantage of large numbers of muons, it becomes possible 
for these experiments 
to study point sources of gamma rays by studying accumulations 
of muons at specific angular positions 
\cite{ap0005180,ap0004379,ap9910549,ap0001111}.
The ability of these experiments to detect a localized source of 
gamma rays by detecting the secondary muons 
depends upon several factors such as a)
the strength, location, duration, and energy 
spectrum of the source; b) the angular resolution, detection area 
and duration of the experiment; and c) the number of muons which reach 
detection level for the relevant region of primary gamma ray energies 
due to the physics of the air blanket covering the surface of 
the earth.  A precise calculation of (c) in the 1 GeV to 10 TeV 
primary gamma ray energy region is the subject of this paper.  

A reliable study of muon production by primary gamma rays is necessary 
in these cases to gauge the sensitivity of the experiments 
to gamma primaries and to
provide information about the expected energy and spatial distribution
of muons. 
Since the threshold energy for detecting the muons has 
been reduced in these experiments, lower muon energies 
and lower primary gamma ray energies than previously investigated 
are studied in this paper. 
Although the total number of muons per gamma ray at ground level 
decreases with lower gamma energies,
electromagnetic showers produced by photons with 
lower energies can produce 
more ground-level muons in total 
due to the steep increase of gamma flux at lower energies (a spectrum 
dN/dE = E$^{-\gamma}$, with $\gamma\,=\,2.4 \pm 0.4$ \cite{EGRET}) 
(assuming the primary gamma energy is well above photopion 
production threshold). 
Therefore, the new calculations presented here include
the primary gamma ray energy range of 
1~GeV~$<$ E$_\gamma < 10$~TeV and secondary muon or electron 
energies above 3~MeV. 
For the results presented in this paper, only primary angles at 0 degrees
from zenith are considered. 

\section{The {\sc FLUKA} program}
\label{fluka.sec}
\subsection{Physics}
{\sc FLUKA}, unlike most Monte Carlo codes
used in cosmic ray research \cite{ALTAI,CORSIKA,AIRES,KASCADE,CHESS,MOCCA,HEMAS} 
is not specialized for this particular field, but is a
multipurpose particle transport program with applications as diverse
as proton and electron accelerator shielding, calorimetry, medical
physics, beam design, high and low energy dosimetry, isotope
production, etc. \cite{Fas97a,Fas97b}. Recently however, it also has 
been used successfully in space and cosmic ray studies \cite{Roe98,Bat98,Bat00}.

In {\sc FLUKA}, different physical models, or event generators, are
responsible for the various aspects (particle type, multiplicity, 
energy and angle) of particle production at 
different energies.
These theoretical models have been directly tested against a large
amount of nuclear experimental data, and have also been indirectly  
validated by comparisons with shower measurements, 
obtained both at accelerators \cite{Fas93,Fer96,Fer97,Aja97} 
and in cosmic ray experiments \cite{Roe98,Pat95,Fer97a}.
In particular, {\sc FLUKA} has been shown to predict 
hadron-generated muon spectra at different heights 
in the atmosphere with good accuracy \cite{Bat99}.

For the present calculations, the following models are relevant 
(more details can be found in \cite{Fer96}):
\begin{itemize}
  \item Hadronic interactions above 4~GeV are simulated according to the 
        Dual Parton Model \cite{Cap94}. A list of improvements to the original 
        Monte Carlo version of the model by Ranft \cite{Ranft} can be found 
        in \cite{Fas97a}. 
  \item A cascade pre-equilibrium model 
        is used for hadronic interactions below 3~GeV. 
        The model includes pion and kaon production. Between 3 and 4~GeV, inelastic 
        hadron collisions are treated according to a resonance-decay model.
  \item The Vector Meson Dominance model is used for photonuclear interactions 
        at energies larger than 4~GeV. The total cross section is based on
        experimental photon-proton and photon-neutron cross sections up to and 
        including HERA energies and scaled to photon-nucleus interaction
        according to Bauer {\em et al.} \cite{Bau78}. Shadowing corrections are 
        based on 
        experimental data. The interaction of vector mesons with the nucleus is 
        handled by the Dual Parton Model.
  \item Delta Resonance excitation (in the framework of the pre-equilibrium model)
        is used for photonuclear interactions below 4~GeV.
\end{itemize}

\begin{figure}[htb]
 \begin{center}
 \mbox{ \epsfig{file=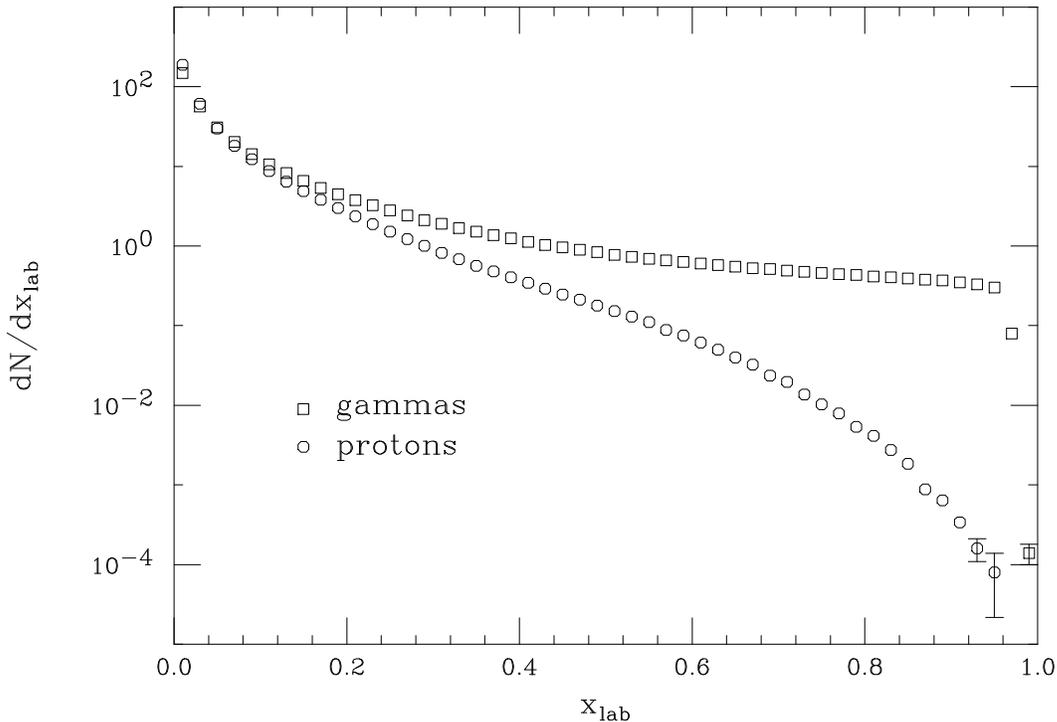,bbllx=0pt,bblly=380pt,bburx=550pt,bbury=0pt,
  width=11.954cm, angle=90} }
 \end{center}
 \vspace*{-2cm}
 \caption{The Feynman-x distribution of charged pions (in the lab frame) for              
100 GeV gamma ray primaries compared with proton primaries of the same energy   
as contained in the FLUKA Monte Carlo generator for the primary                 
interaction in air.}
 \label{feynman}
\end{figure}

To illustrate the critical role played by event generators in predicting the muon 
content of showers, the Feynman-x distribution of charged pions (in the lab frame) 
as calculated by FLUKA is reported in Fig.~\ref{feynman} for both gamma and protons 
at 100~GeV. This figure emphasizes a basic difference between gamma and 
proton-induced showers: the gamma primaries have a larger fraction of high-x 
secondaries than the proton primaries.

If we disregard the lower total cross section for gamma rays to                 
produce mesons (about a factor 100), then gammas appear to be more                       
efficient at producing energetic, forward directed pions, and                  
therefore penetrating muons.                                                
The gamma primaries give rise to higher                                         
energy secondary mesons (on average) and will                                   
thus retain a greater fraction of the primary energy in the                     
numerous subsequent hadronic interactions for those which                       
do not decay.  Although these differences seem rather                           
substantive, the actual differences which can be observed                       
near sea level may be minor. At any rate, in this paper                        
we only investigate the gamma primaries and leave the                           
comparison with hadronic primaries for future consideration.                  

The simulation of the electromagnetic cascade in FLUKA is very accurate,
including the Landau-Pomeranchuk-Migdal effect and a special treatment
of the ``tip'' of the bremsstrahlung spectrum. Electron pairs and
bremsstrahlung are sampled from the proper double differential
energy-angular distributions improving the 
common practice of using average angles.  
In a similar way, the three-dimensional shape of the hadronic cascades
is reproduced in detail by a rigorous sampling of correlated energy
and angles in decay, scattering, and multiple Coulomb scattering.

Bethe-Heitler muon pair production is presently being implemented in
{\sc FLUKA} \cite{Bau99,Roe00}, but was not yet available at the time of the
present calculations; however there is general agreement that this
effect is important only for gamma energies greater than several TeV
and for muon energies larger than 1~TeV \cite{Kud85,Sta85b,Hal86,Ber88,Bha97}. 
Therefore the results presented here should not be affected,
with the possible exception of the highest energy point.
Charm photoproduction, another possible source of muons, is also not available 
in {\sc FLUKA}. According to Berezinskii \cite{Ber87} the corresponding cross section 
at any energy does not exceed a few percent of other muon producing effects.  
Each of these effects, if included, would slightly increase the number 
of muons at ground level above the numbers obtained in this paper.

\subsection{Variance reduction techniques}
Statistical techniques for accelerating convergence of the results
have been used in a few cosmic ray codes, for example 
{\sc MOCCA} \cite{MOCCA}. In {\sc FLUKA}, as in most Monte Carlo codes
used to solve deep penetration shielding problems, several so-called
``biasing'' options are available which allow sampling of events
having a very small probability. Rigorous proofs of the convergence of
results obtained by these techniques to the correct value can be found
in specialized books \cite{Cashwell,Lux}. It is important to remember, 
however, that their use is restricted to the estimation of expectation
values and is not appropriate when studying correlations and fluctuations.

In this study, the use of variance reduction techniques has
proved essential. It is important to realize that the goal was not
to obtain the same results using less computer time, but to include
in the study phase space regions which
would otherwise not be accessible to Monte Carlo techniques.  
Due to the very large number of primary photons, 
some interactions, although extremely rare, may generate events 
having a finite probability to be detected in an experiment. Below
some level of probability {\em per primary photon} the computing time
required to collect a sufficient number of such events by an unbiased 
simulation would become prohibitive.

In the present calculations, the following biasing options have been used:
\begin{description}
  \item{\bf Leading particle biasing:} At each electromagnetic 
  interaction with two particles in the final state (bremsstrahlung,
  pair production, etc.) only one of the two particles is followed,
  with a probability proportional to its energy. Its statistical weight
  is modified so as to conserve total weight. This technique, first
  introduced by A.~Van Ginneken \cite{VanGin}, is very similar -- but not 
  identical -- to the so-called ``thinning algorithm'' of Hillas \cite{MOCCA}.
  \item{\bf Biasing of the mean free path:} For interactions having a very small 
  cross section (in our case photonuclear interactions) the cross section
  is artificially increased by an arbitrary factor chosen by the user 
  (ranging from 10 to 50 in this calculation).  
  The weight of the produced secondaries is reduced so 
  as to conserve probabilities.
  \item{\bf Forced decay:} A similar technique is used to enhance muon production by 
  artificially decreasing the average decay length of charged mesons. 
  Also in this case, since the weights of both the parent meson and of the produced 
  muon are adjusted by the ratio between the actual and the artificial probability,
  all the resulting space, energy and angular distributions are correctly
  reproduced (but with much better statistics in the ranges of interest).
  \item{\bf Importance splitting:} The loss of statistics due to decrease of 
  particle number with depth in the atmosphere is compensated by replacing a 
  Monte Carlo particle with additional identical particles of lower 
  statistical weight when the particle crosses a boundary between
  two regions of different pre-defined statistical importance.
\end{description}

\section{Calculations}
\label{calcul.sec}
The 1999 version of {\sc FLUKA} was used to calculate the muon flux and
kinetic energy spectrum at various altitudes. At sea level and at 222~m 
(the GRAND altitude), muon tracks were scored in 10 concentric rings 
of radii ranging from 10~m to 10~km. Photons of energy between 1~GeV and 
10~TeV were assumed to enter the atmosphere vertically at 80~km and the 
full generated electromagnetic and hadronic showers were followed down to pion
production threshold energy. The mean free path for photonuclear interaction
was artificially shortened to enhance the sampling frequency
by a factor of 10 to 50. As stressed above, the mathematical treatment used
ensures that all results are unaffected, while the variances of the average 
scored quantities (due to the statistical nature of the Monte Carlo technique) 
are reduced to acceptable levels.

Some approximations were adopted which are expected to have a negligible
effect on the results. The atmosphere's geometry was assumed flat and was 
subdivided into 50 layers for the muon calculation, 
each of constant density, in order 
to approximate the exponential character of the earth's air density.
Doubling the number of layers from the 25 used in a previous series of 
calculations did not show any significant difference.
The earth's magnetic field was ignored.  

For comparison purposes, a similar set of calculations was made to
estimate the electron flux at the same positions. All conditions were
identical to the muon calculation 
except the shower energy cutoff was lowered 
from 150 to 3~MeV to accomodate 
the lower energies of interest for the electrons and the atmosphere 
was subdivided into only 25 layers.  The word ``electron'' 
in this paper is used generically; in all cases it means 
e$^{+}$ + e$^{-}$; as well, ``muon'' means $\mu^{+}+\mu^{-}$.  

\section{Results}
\label{results.sec}
\begin{figure}[htb]
 \begin{center}
 \mbox{ \epsfig{file=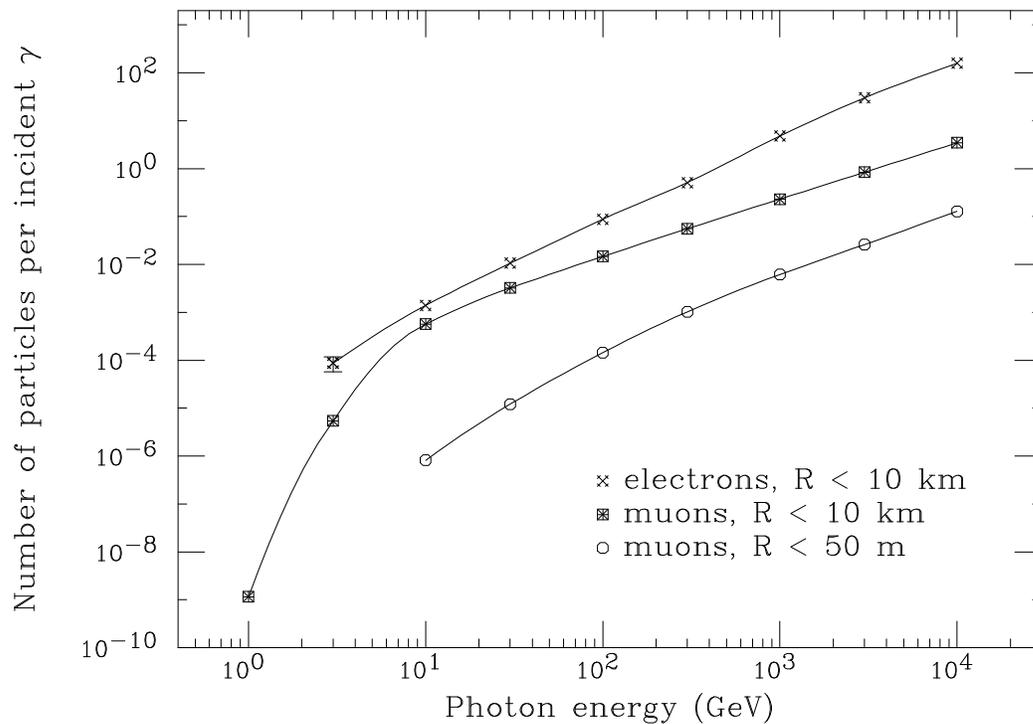,bbllx=0pt,bblly=380pt,bburx=550pt,bbury=0pt,
  width=11.954cm, angle=90} }
 \end{center}
 \vspace*{-2cm}
 \caption{Number of muons and electrons at 222~m a.s.l. as a function of 
 incident primary gamma ray energy.  
The lower points are for secondary muons within 
 a 50~m radius. }
 \label{vsen}
\end{figure}

Fig.~\ref{vsen} shows how the average total number of muons and 
electrons grows with increasing photon energy. A summary of 
calculated data is reported in Table~\ref{oldtab}.  
The entries are for gamma ray primaries vertically 
incident at the top of the earth's atmosphere.  
All error bars in the figures and tables refer to the statistics 
of the Monte Carlo calculation. 

\clearpage

\begin{table}[htb]
\footnotesize
\begin{tabular}{ccccccc}    
\TT $\gamma$ energy & \multicolumn{2}{c}{electrons} & ~~~ & 
    \multicolumn{2}{c}{muons} \\    
\BB (GeV) & 222~m a.s.l. & sea level & ~~~ & 222~m a.s.l. & sea level \\
\hline
\TT\B 10$^4$ & 158  $\pm$ 7   & 121  $\pm$ 6 & ~~~
     & 3.46  $\pm$ 0.05  & 3.31  $\pm$ 0.05 \\
\T\B 3000   & 30   $\pm$ 3   &  23  $\pm$ 3 & ~~~
     & 0.840 $\pm$ 0.004 & 0.807 $\pm$ 0.004 \\
\T\B 1000   & 4.8 $\pm$ 0.7  & 3.7 $\pm$ 0.6 & ~~~
     & 0.2312 $\pm$ 0.0014 & 0.2230 $\pm$ 0.0014 \\
\T\B  300   & 0.51 $\pm$ 0.06 & 0.41 $\pm$ 0.05 & ~~~
     & $(5.55 \pm 0.08) \times 10^{-2}$ & $(5.37 \pm 0.08) \times 10^{-2}$ \\
\T\B  100   & $(8.79 \pm 0.06) \times 10^{-2}$ & $(5.84 \pm 0.04) \times 10^{-2}$ & ~~~
     & $(1.470 \pm 0.008) \times 10^{-2}$ & $(1.424 \pm 0.008) \times 10^{-2}$ \\
\T\B   30   & $(1.06 \pm 0.16)\times 10^{-2}$ & $(8.8 \pm 1.4) \times 10^{-3}$ & ~~~
     & $(3.25 \pm 0.07) \times 10^{-3}$ & $(3.15 \pm 0.07) \times 10^{-3}$ \\
\T\B   10   & $(1.4 \pm 0.3) \times 10^{-3}$ & $(1.0 \pm 0.5) \times 10^{-3}$ & ~~~
     & $(5.66 \pm 0.11) \times 10^{-4}$ & $(5.45 \pm 0.10) \times 10^{-4}$ \\
\T\B    3   & $(8.7 \pm 3.0) \times 10^{-5}$ & $(3.8 \pm 1.6) \times 10^{-5}$ & ~~~
     & $(5.42 \pm 0.09) \times 10^{-6}$ & $(4.32 \pm 0.09) \times 10^{-6}$ \\
\T\BB    1   & ~   & ~  & ~~~~
     & $(1.15 \pm 0.19) \times 10^{-9}$ & $(6.9 \pm 1.4) \times 10^{-10}$ \\
\end{tabular}
\caption{Number of secondary muons and electrons 
per primary gamma ray within a 10~km radius 
at 222 and 0~m above sea level.}
\label{oldtab}
\end{table}

\begin{table}[htb]
\renewcommand{\arraystretch}{1.4}
\footnotesize
\hspace*{-1cm}
\begin{tabular}{cccccccccc}
\TT\B $\gamma$ energy & \multicolumn{5}{c}{muons} \\
\BB       (GeV)       &  5000 m & 4000 m & 3000 m & 2000 m & 1000 m \\
\hline
\TT  10$^4$ & $ 9.59 \pm 0.08 $                   & $ 8.13 \pm 0.07 $                  &	
              $ 6.56 \pm 0.07 $                   & $ 5.16 \pm 0.06 $                  & 
              $ 4.09 \pm 0.05 $                   \\
     3000   & $ 2.420 \pm 0.007$                  & $ 1.966 \pm 0.007$                 &	
              $ 1.547 \pm 0.007$                  & $ 1.218 \pm 0.006$                 & 
              $ 0.978 \pm 0.005$                  \\
     1000   & $0.6661 \pm 0.0019$                 & $0.525 \pm 0.002$                  &	
              $0.4105 \pm 0.0017$                 & $0.325 \pm 0.002$                  & 
              $0.2659 \pm 0.0016$                 \\
      300   & $0.1505 \pm 0.0014$                 & $ 0.1175  \pm 0.0012$              &	
              $ (9.29 \pm  0.11) \times 10^{-2}$  & $ (7.54 \pm  0.10) \times 10^{-2}$ & 
              $ (6.28 \pm  0.09) \times 10^{-2}$  \\
      100   & $(3.727 \pm 0.017) \times 10^{-2}$  & $(2.919 \pm 0.015) \times 10^{-2}$ &	
              $(2.349 \pm 0.011) \times 10^{-2}$  & $(1.939 \pm 0.010) \times 10^{-2}$ & 
              $(1.647 \pm 0.008) \times 10^{-2}$  \\
       30   & $ (7.75 \pm 0.17) \times 10^{-3}$   & $ (6.16 \pm 0.13) \times 10^{-3}$  &	
              $ (5.03 \pm 0.10) \times 10^{-3}$   & $ (4.25 \pm 0.08) \times 10^{-3}$  & 
              $ (3.65 \pm 0.08) \times 10^{-3}$   \\
       10   & $(1.46 \pm 0.03) \times 10^{-3}$    & $(1.16 \pm 0.02) \times 10^{-3}$   &	
              $(9.4 \pm 0.2) \times 10^{-4}$      & $(7.71 \pm 0.15) \times 10^{-4}$   & 
              $(6.43 \pm 0.13) \times 10^{-4}$    \\
        3   & $ (1.076 \pm 0.010) \times 10^{-4}$ & $ (6.86 \pm 0.08) \times 10^{-5}$  &	
              $ (4.15 \pm 0.05) \times 10^{-5}$   & $ (2.31 \pm 0.04) \times 10^{-5}$  & 
              $ (1.102 \pm 0.019) \times 10^{-5}$ \\
\BB     1   & $ (1.26 \pm 0.15) \times 10^{-6}$   & $ (2.2 \pm 0.3) \times 10^{-7}$    & 
              $ (7 \pm  3) \times 10^{-8}$        & $ (5.6 \pm 0.6) \times 10^{-9}$    & 
              $ (2.8 \pm 0.4) \times 10^{-9}$     \\
\end{tabular}
 \caption{Number of secondary muons per incident gamma ray within a 
 10~km radius at different heights above sea level.}
 \label{newtab}
\end{table}

\begin{figure}[htb]
 \begin{center}
 \mbox{ \epsfig{file=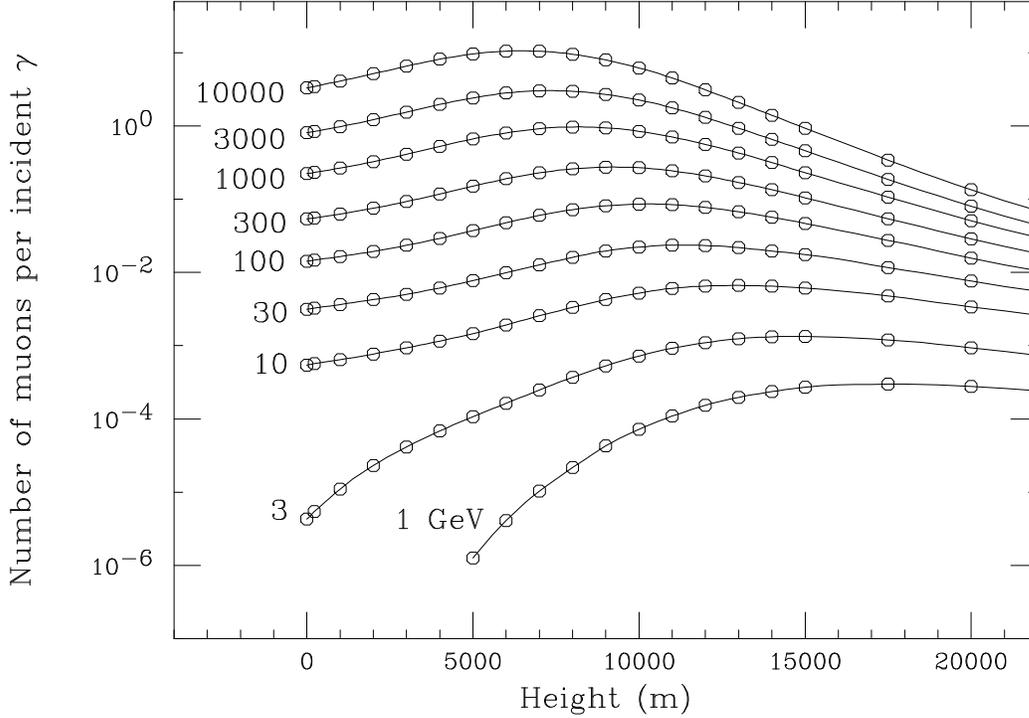,bbllx=0pt,bblly=380pt,bburx=550pt,bbury=0pt,
  width=11.954cm, angle=90} }
 \end{center}
 \vspace*{-2cm}
 \caption{Average number of secondary muons per incident gamma ray within a 
 10~km radius at different heights above sea level.}
 \label{heights}
\end{figure}

Fig.~\ref{heights} shows how the total number of muons depends upon height 
above sea level for nine gamma ray primary energies 
and elevations from 0 to 20~km. 
The corresponding numerical data are contained in Table~\ref{newtab} for heights 
from 1~km to 5~km.
One could interpolate among the values in Table~\ref{newtab} to 
estimate the expected number of muons per gamma primary for
ground-based detectors at intermediate heights.
As expected, the most probable height for each primary energy 
gradually shifts toward lower heights as the primary energy increases.

\begin{figure}[htb]
 \begin{center}
 \mbox{ \epsfig{file=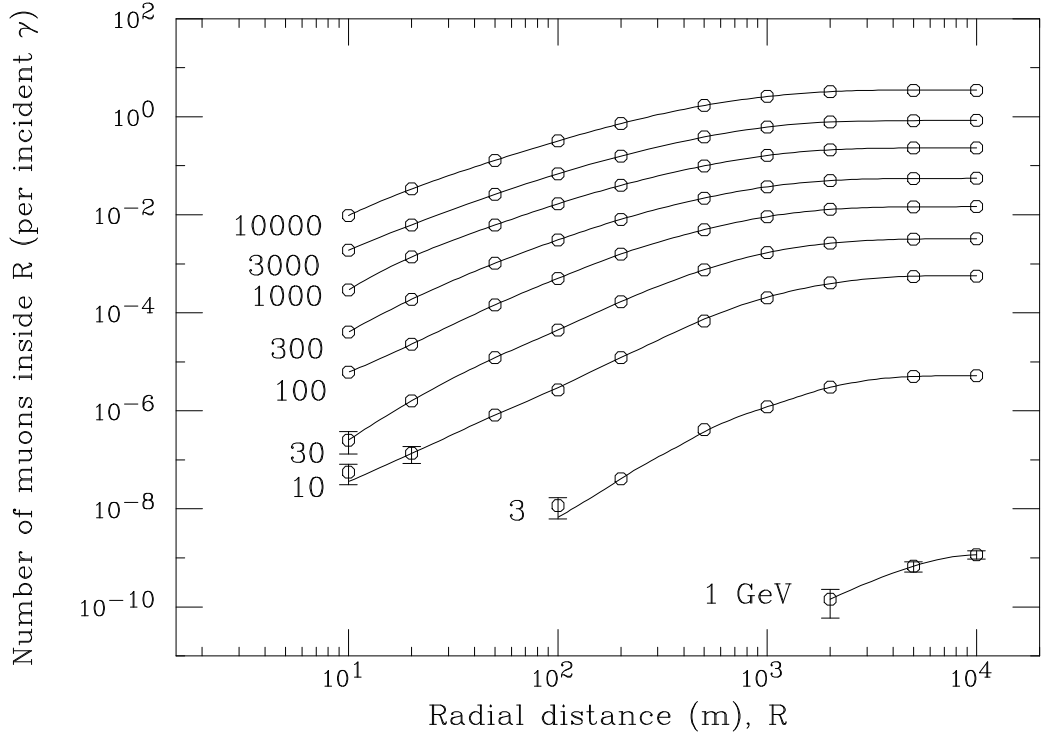,bbllx=0pt,bblly=380pt,bburx=550pt,bbury=0pt,
  width=11.954cm, angle=90} }
 \end{center}
 \vspace*{-2cm}
 \caption{Number of muons per incident photon 
  within the stated radial distance.  
  Nine incident primary gamma ray energies are shown.}
 \label{radial}
\end{figure}

\pagebreak
The dependence on radial distance is shown in Fig.~\ref{radial}. The total 
number
of muons within a given radius is plotted as a function of radius.
For photon energies $\ge$ 300~GeV, 
the circle containing half of all the muons reaching 
sea level has a radius $\le$ 700~m,
increasing to $\ge$ 1600~m for photon energies $\le$ 3~GeV.

\begin{figure}[htb]
 \begin{center}
 \mbox{ \epsfig{file=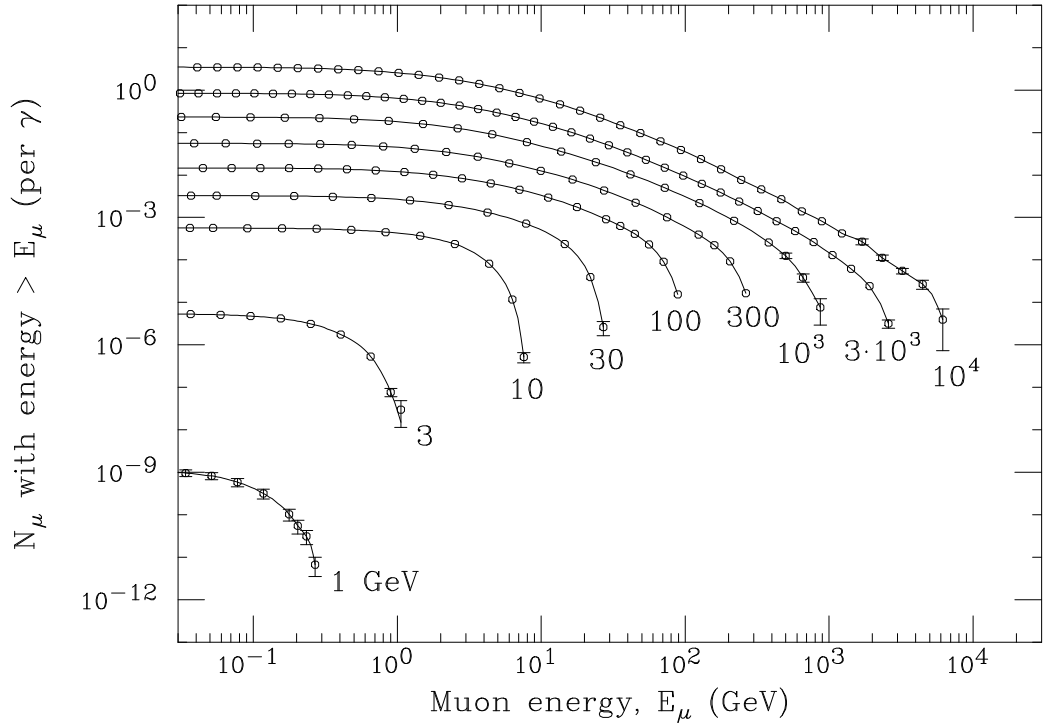,bbllx=0pt,bblly=380pt,bburx=550pt,bbury=0pt,
  width=11.954cm, angle=90} }
 \end{center}
 \vspace*{-2cm}
 \caption{Integral spectra of muon kinetic energy at radial distances 
 $<$~10~km from the shower center, at 222~m a.s.l.}
 \label{spectra}
\end{figure}

A representative example of the numerous spectra which have been calculated 
is shown in Fig.~\ref{spectra} which 
shows nine different primary photon energies from 1~GeV to 10~TeV 
and gives the muon energy spectra at 222 m above sea level for those muons 
at distances $<$\,10~km (essentially all of them).  
For a given primary gamma ray energy the points and their error bars 
are highly correlated due to the summations involved in getting the 
integrals presented.   

\vspace*{-1.4cm}

To estimate the total number of muons reaching detection level, the calculated 
data should be folded with the incident primary photon spectrum.  
Assuming a differential energy dependence of E$^{-2.41}$ \cite{EGRET}, 
it is found 
that the largest number of muons comes from primary energies of 
27~GeV.  If the spectrum were softer (spectral index, $\gamma\,=\,2.77$), 
the corresponding 
primary energy would drop to 12~GeV; if the spectrum were harder 
($\gamma\,=\,2.05$), the muon number would continue to increase as the 
primary energy increases and would depend on where the spectrum 
softens or cuts off.  

\section{Conclusions}	
These reported results are from the Monte Carlo program {\sc FLUKA} 
which contains event generators built on the accurate 
experimental data available in the the energy region of interest.
The use of biasing techniques enabled the calculation of events originated 
by low energy primaries with muon
probabilities as low as $\approx 10^{-9}$ per incident gamma ray,
which are important since the 
primary cosmic ray flux rises rapidly with decreasing primary energy.  
This {\sc FLUKA}-based Monte Carlo calculation is believed to be quite 
accurate in this energy region.
As such, the results give a good 
estimate of the number of muons expected from gamma ray primaries 
in this energy region and help overcome the stigma that muons are only 
anticoincidence signals for gamma ray primaries.  
Instead, with the high statistics of the experimental results that 
are becoming available in this 1~GeV to 10~TeV primary energy region, 
a statistically significant excess accumulation of muons at a specific 
angular direction becomes a positive signature 
for gamma ray initiated hadronic showers and is thus a tool to fill 
the energy gap between satellite/balloon experiments and Cerenkov arrays.  
Examples of this use are contained in references \cite{ap0001111,grb}.

\vspace*{8mm}\noindent
Thanks to S.~Roesler, M.~Dunford, and T.~Bowen for their help with 
this paper.  
Part of this work was supported by the Department of Energy under contract 
DE-AC03-76SF00515. 
Project GRAND is funded through grants from the University of
Notre Dame and private donations. \\


\end{document}